\documentclass[pra,aps,nopacs,showkeys,twocolumn,twoside]{revtex4}

\usepackage{graphicx,epic,eepic,epsfig,amsmath,latexsym,amssymb,verbatim}

\usepackage{theorem}
\newtheorem{definition}{Definition}
\newtheorem{proposition}[definition]{Proposition}
\newtheorem{lemma}[definition]{Lemma}

\newtheorem{theorem}[definition]{Theorem}
\newtheorem{corollary}[definition]{Corollary}

\newtheorem{expl}[definition]{Example}
\newtheorem{remark}[definition]{Remark}

\def\squareforqed{\hbox{\rlap{$\sqcap$}$\sqcup$}}
\def\qed{\ifmmode\squareforqed\else{\unskip\nobreak\hfil
\penalty50\hskip1em\null\nobreak\hfil\squareforqed
\parfillskip=0pt\finalhyphendemerits=0\endgraf}\fi}
\def\endenv{\ifmmode\;\else{\unskip\nobreak\hfil
\penalty50\hskip1em\null\nobreak\hfil\;
\parfillskip=0pt\finalhyphendemerits=0\endgraf}\fi}

\newlength{\blank}
\newlength{\equalsign}
\newenvironment{beweis}[1][{\hspace{-\blank}}]{{\noindent\emph{Proof~{#1}.\ }}}{\hfill\qed\vskip 0.5\baselineskip}

\mathchardef\ordinarycolon\mathcode`\:
\mathcode`\:=\string"8000
\def\vcentcolon{\mathrel{\mathop\ordinarycolon}}
\begingroup \catcode`\:=\active
  \lowercase{\endgroup
  \let :\vcentcolon
  }

\newcommand{\nc}{\newcommand}
\nc{\rnc}{\renewcommand}
\nc{\beq}{\begin{equation}}
\nc{\eeq}{{\end{equation}}}
\nc{\beqa}{\begin{eqnarray}}
\nc{\eeqa}{\end{eqnarray}}
\nc{\lbar}[1]{\overline{#1}}
\nc{\bra}[1]{\langle#1|}
\nc{\ket}[1]{|#1\rangle}
\nc{\ketbra}[2]{|#1\rangle\!\langle#2|}
\nc{\braket}[2]{\langle#1|#2\rangle}
\nc{\proj}[1]{| #1\rangle\!\langle #1 |}
\nc{\avg}[1]{\langle#1\rangle}
\rnc{\max}{\operatorname{max}}
\nc{\Rank}{\operatorname{Rank}}
\nc{\smfrac}[2]{\mbox{$\frac{#1}{#2}$}}
\nc{\tr}{\operatorname{Tr}}
\nc{\ox}{\otimes}
\nc{\dg}{\dagger}
\nc{\dn}{\downarrow}
\nc{\cA}{{\cal A}}
\nc{\cB}{{\cal B}}
\nc{\cC}{{\cal C}}
\nc{\cD}{{\cal D}}
\nc{\cE}{{\cal E}}
\nc{\cF}{{\cal F}}
\nc{\cG}{{\cal G}}
\nc{\cH}{{\cal H}}
\nc{\cI}{{\cal I}}
\nc{\cJ}{{\cal J}}
\nc{\cK}{{\cal K}}
\nc{\cL}{{\cal L}}
\nc{\cM}{{\cal M}}
\nc{\cO}{{\cal O}}
\nc{\cP}{{\cal P}}
\nc{\cR}{{\cal R}}
\nc{\cS}{{\cal S}}
\nc{\cT}{{\cal T}}
\nc{\cX}{{\cal X}}
\nc{\cZ}{{\cal Z}}
\nc{\csupp}{{\operatorname{csupp}}}
\nc{\qsupp}{{\operatorname{qsupp}}}
\nc{\var}{\operatorname{var}}
\nc{\rar}{\rightarrow}
\nc{\lrar}{\longrightarrow}
\nc{\polylog}{\operatorname{polylog}}
\nc{\1}{{\openone}}
\nc{\GHZ}{{\Gamma}}
\nc{\EPR}{{\Phi_2}}

\def\a{\alpha}

\def\d{\delta}
\def\e{\epsilon}

\def\ph{\varphi}

\def\G{\Gamma}

\nc{\RR}{{{\mathbb R}}}
\nc{\CC}{{{\mathbb C}}}
\nc{\FF}{{{\mathbb F}}}
\nc{\NN}{{{\mathbb N}}}
\nc{\ZZ}{{{\mathbb Z}}}
\nc{\PP}{{{\mathbb P}}}
\nc{\QQ}{{{\mathbb Q}}}
\nc{\UU}{{{\mathbb U}}}
\nc{\EE}{{{\mathbb E}}}
\nc{\id}{{\operatorname{id}}}
\nc{\Span}{{\operatorname{span}}}

\nc{\be}{\begin{equation}}
\nc{\ee}{{\end{equation}}}
\nc{\bea}{\begin{eqnarray}}
\nc{\eea}{\end{eqnarray}}
\nc{\<}{\langle}
\rnc{\>}{\rangle}
\nc{\Hom}[2]{\mbox{Hom}(\CC^{#1},\CC^{#2})}
\nc{\rU}{\mbox{U}}

\nc{\ob}[1]{#1}

\begin{document}

\title{On environment-assisted capacities of quantum channels}

\author{Andreas Winter}
\affiliation{Department of Mathematics, University of Bristol,\protect\\
             University Walk, Bristol BS8 1TW, U.K.\protect\\
             Email: {\tt a.j.winter@bris.ac.uk}}

\date{5th July 2005}

\begin{abstract}
  Following initial work by Gregoratti and Werner [J. Mod. Optics,
  vol. 50, no. 6\&7, pp. 913-933, 2003 and {\tt quant-ph/0403092}]
  and Hayden and King [{\tt quant-ph/0409026}],
  we study the problem of the capacity of a quantum channel
  assisted by a ``friendly (channel) environment'' that
  can locally measure and communicate classical messages to
  the receiver.

  Previous work [{\tt quant-ph/0505038}] has yielded a
  capacity formula for the quantum capacity under this kind of
  help from the environment. Here we study the problem of
  the environment-assisted classical capacity, which exhibits
  a somewhat richer structure (at least, it seems to be the harder
  problem). There are several, presumably inequivalent, models
  of the permitted local operations and classical
  communications between receiver and environment: one-way,
  arbitrary, separable and PPT POVMs.
  In all these models, the task of decoding a message amounts to
  discriminating a set of possibly entangled states between
  the two receivers, by a class of operations under some sort of
  locality constraint.

  After introducing the operational capacities outlined above,
  we show that a lower bound on the environment-assisted classical capacity
  is always half the logarithm of the input space dimension.
  Then we develop a few techniques to prove the existence of
  channels which meet this lower bound up to terms of
  much smaller order, even when PPT decoding measurements
  are allowed (assuming a certain superadditivity conjecture).
\end{abstract}

\keywords{entanglement of assistance, quantum error correction, feedback control,
          LOCC discrimination, PPT discrimination, additivity conjecture}

\maketitle

\section{Introduction and background}
\label{sec:intro}
A noisy quantum channel is modelled universally as a completely
positive and trace preserving (cptp) map
$${\cal N}: A \longrightarrow B$$
between the algebras of observables $A = {\cal B}({\cal H}_A)$ and
$B = {\cal B}({\cal H}_B)$, which we assume to be finite-dimensional
throughout.
It can always be presented as an isometry
$$U: {\cal H}_A \longrightarrow {\cal H}_B \otimes {\cal H}_C,$$
followed by the partial trace map $\tr_C: B \otimes C \longrightarrow B$.
This is the content of Stinespring's theorem~\cite{stinespring},
which also informs us that the isometry is unique up to unitaries
on ${\cal H}_C$, which system is usually called the ``channel
environment''. This means that associated with ${\cal N}$
there is a canonical ``dual channel''
$$\overline{{\cal N}}: A \longrightarrow C,$$
defined as the above isometry $U$ followed by the other
partial trace map $\tr_B: B \otimes C \longrightarrow C$.

We shall here be interested in information 
transmission from A(lice) to B(ob) the channel,
when assisted by the environment (Charlie), specifically
in its information theoretic version of obtaining
bounds on asymptotic rates.

An important case is where Bob and Charlie are allowed arbitrary 
local operations and classical communication (LOCC) to extract
a classical encoded by Alice: the signals are then states
$\ket{\phi} \in {\cal S} := U{\cal H}_A \subset {\cal H}_B \otimes {\cal H}_C$
in the image subspace ${\cal S}$ in the joint Bob-Charlie system,
and the decoding task is to discriminate a set of these
entangled states by measurements restricted by the LOCC constraint.

There are various important restrictions and relaxations of this
model: we may insist on one-way LOCC from Charlie to Bob,
or one-way LOCC from Bob to Charlie, or we may allow
arbitrary LOCC. Since the class of all unrestricted LOCC
operations, or even measurements, is notoriously hard to
characterise, it is convenient for mathematical analysis to go
to the larger class of separable POVMs, i.e. measurements
whose POVM operators are sums of positive product operators,
or in the even wider class of PPT (positive partial transpose)
operators, as pioneered in Rains' work~\cite{rains}:
for $M = \sum_{ij,kl} M_{ij,kl}\ket{ij}\bra{kl}$
(in an arbitrarily fixed basis), we demand
$M^\Gamma := \sum_{ij,kl} M_{ij,kl}\ket{il}\bra{kj} \geq 0$.
It has been noticed before~\cite{nonlocal:w-o-ent} that there
are indeed separable POVMs which are not LOCC, and it is quite
easy to see that there exist PPT POVMs which are not separable.
Discriminating states via LOCC has become quite a large field,
and here we can only collect a few pointers to the most
significant papers (and references therein):
Walgate et al.~\cite{walgate:etal},
Walgate and Hardy~\cite{walgate:hardy},
Bennett et al.~\cite{nonlocal:w-o-ent},
Chefles~\cite{chefles},
Ghosh et al.~\cite{ghosh:etal}
and the more recent investigations by
Badzi\c{a}g et al.~and Ghosh et al.~\cite{badziag:etal},
Nathanson~\cite{nathanson}
and Owari and Hayashi~\cite{owari:hayashi}.

The structure of the paper is as follows: in the next section we
will consider the problem of environment-assisted quantum 
capacity, and revisit the recently obtained capacity formula~\cite{EoA:oho}.
Then, in section~\ref{sec:C-ass} we introduce the relevant notions
of environment-assisted transmission codes and
the corresponding capacities, and present various lower
bounds. Section~\ref{sec:upperbound} quotes a nontrivial upper bound
on the LOCC-assisted classical capacity from~\cite{badziag:etal},
and presents an extension of it adapted to the more general
class of PPT POVMs.
Then, in section~\ref{sec:example} we exhibit a class of examples
for which the PPT-decoded classical capacity almost meets the general
lower bound derived earlier, after which we conclude, highlighting a
few open questions. An appendix quotes some technical results
from the literature.

\section{Quantum capacity with classical helper in the environment}
\label{sec:channel+helper}
Gregoratti and Werner~\cite{lost:found} consider the channel
model with helper in the environment, as outlined in the introductory
section: an isometry $U$ from Alice's input system $A$ to
the combination of Bob's output system $B$ and the environment $C$.
Assume that the environment system may be measured and the
classical results of the observation be forwarded to Bob --- attempting
to help him in error correcting quantum information sent from Alice.

\begin{figure}[ht]
  \centering
  \includegraphics[width=7.5cm]{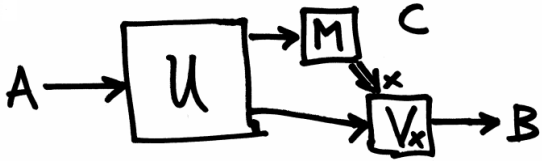}
  \caption{Alice prepares an input to (many copies of) the isometry
           $U$, which gives part of the state to Bob and part to Charlie.
           The latter measures a POVM $M$ on his system and classically
           communicates his result $x$ to Bob, who executes a unitary $V_x$
           depending on Charlie's message to recover Alice's input state.}
  \label{fig:lost+found}
\end{figure}

We are interested, for this scenario, in the quantum transmission capacity
from Alice to Bob, in the asymptotic limit of block coded
information (and collectively measured environment).
The setup is illustrated in figure~\ref{fig:lost+found}.
Formally, an environment-assisted quantum code
(on block length $n$) is defined to consist of an encoder (cptp map)
${\cal E}: {\cal B}({\cal H}) \longrightarrow {\cal B}({\cal H}_A^{\otimes n})$,
a POVM $(M_x)_x$ on ${\cal H}_C^{\otimes n}$ and cptp maps
${\cal R}_x: {\cal B}({\cal H}_B^{\otimes n}) \longrightarrow {\cal B}({\cal H})$;
the idea is that Alice uses ${\cal E}$ to encode the quantum states
she wants to send, Charlie performs the POVM $(M_x)_x$ and sends $x$ on to
Bob who acts with the map ${\cal R}_x$ on the channel output.
The overall dynamics ${\cal M}:{\cal B}({\cal H}) \rightarrow {\cal B}({\cal H})$
of this setup is
$${\cal M}(\psi) = \sum_x {\cal R}_x\Bigl(
                            \tr_{C^n}\bigl[ (U^{\otimes n} {\cal E}(\psi) U^{*\otimes n})
                                               (\1_B^{\otimes n}\otimes M_x) \bigr] \Bigr),$$
and we say that the code has error $\e$ if for all $\ket{\psi}\in{\cal H}$,
$\frac{1}{2}\| \psi - {\cal M}(\psi) \|_1 \leq \e$. Incidentally,
we will follow the convention of denoting a state vector
always as a ket: e.g.~$\ket{\psi}$, but its pure state
density operator as $\psi = \proj{\psi}$.
Denoting by $M(n,\e)$ the
largest $\dim{\cal H}$ such that an environment-assisted quantum code on block
length $n$ and with error $\e$ exists, we can define the
(optimistic/pessimistic) \emph{environment-assisted quantum capacity} as
$$\inf_{\e>0} \left( \limsup_{n\rightarrow\infty}/\liminf_{n\rightarrow\infty}
                                                \frac{1}{n} \log M(n,\e) \right),$$
respectively.

In previous work by Smolin et al.~\cite{EoA:oho}, the
following result was proved:

\begin{theorem}[\cite{EoA:oho}, Thm.~8]
  \label{thm:lost+found}
  The environment-assisted quantum capacity of the noisy
  channel ${\cal N}: A \longrightarrow B$ (both optimistically and pessimistically)
  is given by
  $$Q_A({\cal N}) = \max_\rho \, \min\bigl\{ S(\rho),S\bigl({\cal N}(\rho)\bigr) \bigr\}.$$
  The same capacity is obtained allowing unlimited LOCC between
  Alice, Bob and Charlie. \qed
\end{theorem}

In particular, if the channel ${\cal N}$ is unital (i.e., preserving the
identity) and $d_B \geq d_A$, then $Q_A({\cal N}) = \log d_A$.
In other words, all of the channel's input bandwidth can be
corrected by looking at the environment.
Since Gregoratti and Werner~\cite{lost:found} have shown that perfect
correction is possible if and only if the channel is a random mixture
of isometries, this can be understood as saying that a unital channel
${\cal N}$ becomes, in the limit of many independent copies, almost a mixture
of isometries. See~\cite{EoA:oho} for a deeper discussion of this point.

\section{Environment- and LOCC-assisted\protect\\ classical capacities}
\label{sec:C-ass}
The isometry $U$ identifies ${\cal H}_A$ with the subspace
${\cal S} = U{\cal H}_A \subset {\cal H}_B \otimes {\cal H}_C$, so
we can define a classical transmission code of blocklength $n$
as follows: it is a family $(\varphi_i,M_i)_{i=1}^N$
of pure states $\ket{\varphi_i} \in {\cal S}^{\otimes n}$
and a POVM $(M_i)_{i=1}^N$ on ${\cal H}_B^{\otimes n} \otimes {\cal H}_C^{\otimes n}$.
We say that the code has error probability $\epsilon$
if for all $i$, $\tr(\varphi_i M_i) \geq 1-\epsilon$.
Some authors prefer the average error probability $\overline{\e}$ as
opposed to the maximal error we consider here: as in Shannon~\cite{shannon}
it is easy to see that by expurgating the large-error signals,
one can sacrifice a fraction $1/a$ of the messages and retain
a set with maximal error $a\overline{\e}$.
Furthermore, with respect to the bipartition Bob-Charlie,
we call the code
\begin{itemize}
  \item \emph{environment-assisted} if the POVM is implemented
    by one-way LOCC from Charlie to Bob;
  \item \emph{environment-assisting} if the POVM is implemented
    by one-way LOCC from Bob to Charlie;
  \item \emph{LOCC-assisted} if the POVM is implemented by
    some LOCC protocol;
  \item \emph{separable-decoding} if the POVM is separable;
  \item \emph{PPT-decoding} if the POVM consists of PPT operators.
\end{itemize}
The largest $N$ such that a code of blocklength $n$ and error probability
$\e$ exists under the above restrictions, are denoted
$N_A^\rightarrow(n,\e)$, $N_A^\leftarrow(n,\e)$, $N_A^\leftrightarrow(n,\e)$, 
$N_A^{\rm sep}(n,\e)$, $N_A^{\rm ppt}(n,\e)$, respectively.

Now we can define capacities in the usual way (cf.~also the previous
section): for example, the (one-way) environment-assisted classical
capacity $C_A^\rightarrow({\cal N})$ is given by
$$\inf_{\e>0} \left( \limsup_{n\rightarrow\infty} \frac{1}{n} \log N_A^\rightarrow(n,\e)
                \right)$$
in the optimistic version, and by
$$\inf_{\e>0} \left( \liminf_{n\rightarrow\infty} \frac{1}{n} \log N_A^\rightarrow(n,\e)
                \right)$$
in the pessimistic version, and likewise for
$C_A^\leftarrow({\cal N}) = C_A^\rightarrow(\overline{{\cal N}})$,
$C_A^\leftrightarrow({\cal N})$, $C_A^{\rm sep}({\cal N})$ and
$C_A^{\rm ppt}({\cal N})$.
We will not introduce special symbols for to distinguish optimistic and
pessimistic capacities, but in this paper follow the convention that
lower bounds on capacities are always proved pessimistically,
and upper bounds optimistically.

Note that the models $\leftrightarrow$, ${\rm sep}$ and ${\rm ppt}$ are
symmetric between Bob and Charlie; hence we denote,
e.g.~$C_A^{\rm ppt}({\cal N}) = C_A^{\rm ppt}(U) = C_A^{\rm ppt}({\cal S})$,
etc.

It should be obvious how to make the connection with the
previously introduced capacity notions~\cite{lost:found,hayden:king}:
for example, it is quite easy to see, using the well-known Fano
inequality, that our capacity $C_A^\rightarrow({\cal N})$ is
the regularised ``corrected Shannon capacity'' $C_{\rm corr}({\cal N})$
of Hayden and King~\cite{hayden:king}:
$$C_A^\rightarrow({\cal N})
   = \lim_{n\rightarrow\infty} \frac{1}{n} C_{\rm corr}\bigl( {\cal N}^{\otimes n} \bigr).$$
Clearly, we have the chain of inequalities
$$Q_A({\cal N}) \leq C_A^\rightarrow({\cal N})
                \leq C_A^\leftrightarrow({\cal N})
                \leq C_A^{\rm ppt}({\cal N}) \leq \log d_A,$$
because every code to the left gives rise to or is itself immediately a
code to the further right, and on the far right we have the input bandwidth,
which is the capacity if B and C are permitted arbitrary joint operations.

%
%

For the formulation of the following general lower
bound on $C_A^\rightarrow({\cal N})$, let us introduce some
notation: for a state $\rho$ on Alice's input system
$A$, consider a generic purification $\phi$ on $A\otimes A$,
and let $\ket{\psi}_{ABC} = (\1_A\otimes U)\ket{\phi}_{AA}$.
Then denote the entropies of the reduced states of $\psi$
by referring to the subsystem(s) to which we restrict the
state: e.g.~$S(A) = S(\rho)$, $S(B) = S\bigl({\cal N}(\rho)\bigr)$,
$S(AB) = S\bigl( \tr_C\psi \bigr) = S(C) = S\bigl( \overline{{\cal N}}(\rho) \bigr)$,
etc. The quantum mutual
information is formally defined as
$$I(A:B) = S(A)+S(B)-S(AB) = S(A)+S(B)-S(C).$$
For another state $\rho'$, we refer to the corresponding entropies
by affixing primes: $S(A')$, $S(C')$, etc.

For example, theorem~\ref{thm:lost+found} implies that
$C_A^\rightarrow({\cal N}) \geq \min\{ S(A),S(B) \}$ since one can
always encode one bit in each qubit that is faithfully transmitted.
Of course, we get by the same token
$C_A^\leftarrow({\cal N}) = C_A^\rightarrow(\overline{{\cal N}})
                          \geq \min\{ S(A),S(C) \}$.
By subadditivity of the entropy, $S(A) = S(BC) \leq S(B)+S(C)$,
so the larger of $S(B)$ and $S(C)$ is at least $\frac{1}{2}S(A)$.
Hence,
\begin{equation}
  \label{eq:LOCC:lower}
  C_A^\leftrightarrow(U)
           \geq \max\bigl\{ C_A^\rightarrow({\cal N}), C_A^\leftarrow({\cal N}) \bigr\}
                                                                \geq \frac{1}{2}\log d_A.
\end{equation}
Note that in general, by the above,
\begin{equation*}\begin{split}
  C_A^\rightarrow({\cal N}) &\geq \min\bigl\{ S(A),S(B) \bigr\} \\
                            &=    \frac{1}{2}\bigl[ S(A)+S(B)-|S(A)-S(B)| \bigr] \\
                            &\geq \frac{1}{2}\bigl[ S(A)+S(B)-S(C) \bigr]
                             =    \frac{1}{2}I(A:B),
\end{split}\end{equation*}
the last line by the triangle inequality.
Also, $\frac{1}{2}I(A:B) \geq \frac{1}{2}S(A)$ if $S(B) \geq S(C)$.
And, if $S(A) \leq S(B)$, even
$C_A^\rightarrow({\cal N}) \geq S(A) \geq \frac{1}{2}S(A)$.

We shall now prove that also
in the remaining case, $S(B) < S(C)$ and $S(B) < S(A)$,
this lower bound holds, thus improving on eq.~(\ref{eq:LOCC:lower}),
in fact something a bit better.
We will use the following recent result:
\begin{lemma}[State merging~\cite{HOW}]
  \label{lemma:merging}
  Let $\ket{\psi}_{ABC}$ be a tripartite pure state with
  $S(A) = S(BC) < S(B)$. Then, for all $\e>0$ and all large enough $n$,
  there exists a measurement $(M_x)_x$ on $C^n$ and a family of
  isometries
  $V_x:{\cal H}_C^{\otimes n} \rightarrow {\cal H}_B^{\otimes n}\otimes{\cal H}_C^{\otimes n}$
  such that
  $$\left\| \psi^{\otimes n}
              - \sum_x V_x
                       \tr_{C^n}\bigl[ \psi^{\otimes n}(\1_{A^n B^n}\otimes M_x) \bigr]
                       V_x^*
    \right\|_1                                            \leq \e.$$
  If $S(A) \geq S(B)$, first sharing of $n\bigl( S(A)-S(B) \bigr) + o(n)$ ebits
  of entanglement creates a state which satisfies the above condition.
  \qed
\end{lemma}

The protocol is based on time sharing between a block of length $k$
that is used to communicate $\sim k \bigl( S(C)-S(B) \bigr)$
bits from Alice to Charlie (who hands on the decoded message to Bob)
and leaving $\sim k S(B)$ ebits between Bob and Charlie;
and a block of length $\ell$ where Alice encodes
$S(A')$ bits into a pure-state ensemble for $\rho'$ (and we assume
$S(A') > S(B')$ here), Charlie merges his state with Bob's
(lemma~\ref{lemma:merging}),
using the previously extracted entanglement, so that Bob
can read Alice's message, of $\sim \ell\, S(A')$ bits.
On the first block we use random quantum coding for the channel
$\overline{{\cal N}}$, see~\cite{devetak:Q,HOW}
which justifies the transmission rate (of quantum information
but we use an orthogonal basis in the code space to transmit
classical information), and the remaining entanglement:
see~\cite{devetak:Q} for a description of the decoding
via a unitary in Charlie's system, which separates the Alice's
quantum message from the remaining entanglement.
Per copy of the state, merging requires $S(A')-S(B')$ ebits and classical
communication from Charlie to Bob~\cite{HOW}, so we must have
$k S(B) \sim \ell\bigl( S(A')-S(B') \bigr)$.
The rate is now the total information transmitted,
$\sim k \bigl( S(C)-S(B) \bigr) + \ell\, S(A')$ bits, divided
by the blocklength $k+\ell$. Thus we have proved:
\begin{theorem}
  \label{thm:C-ass:lower}
  For the (one-way) environment-assisted classical capacity of the
  channel ${\cal N}$, and any input state $\rho$,
  \begin{equation}
    \label{eq:C-ass:lower-1}
    C_A^\rightarrow({\cal N}) \geq \begin{cases}
                                     S(A)              & \text{if }S(A)\leq S(B), \\
                                     \frac{1}{2}I(A:B) & \text{in general}.
                                   \end{cases}
  \end{equation}
  For input states $\rho$ such that $S(B) < S(C)$, and $\rho'$
  such that $S(B') < S(A')$:
  \begin{equation}
    \label{eq:C-ass:lower-2}
    C_A^\rightarrow({\cal N})
            \geq \frac{S(C)-S(B)+\frac{S(B)}{S(A')-S(B')}S(A')}{1+\frac{S(B)}{S(A')-S(B')}},
  \end{equation}
  so that for $\rho=\rho'$ with $S(B) < S(C)$ and $S(B) < S(A)$,
  \begin{equation}
    \label{eq:C-ass:lower-3}
    C_A^\rightarrow({\cal N}) \geq \left[ 1-\frac{S(B)}{S(A)} \right] S(C)
                                                  + \frac{S(B)}{S(A)} S(B).
  \end{equation}
  \qed
\end{theorem}

\begin{corollary}
  \label{cor:C-ass:lower}
  For every channel ${\cal N}$ and input state $\rho$,
  \begin{equation}
    \label{eq:C-ass:lower}
    C_A^\rightarrow({\cal N}) \geq \max\left\{ \frac{1}{2}I(A:B),\frac{1}{2}S(A) \right\},
  \end{equation}
  which for the maximally mixed input state $\rho = \frac{1}{d_A}\1$ gives
  that for all channels,
  \begin{equation}
    \label{eq:C-ass:lower2}
    C_A^\rightarrow({\cal N}) \geq \frac{1}{2}\log d_A.
  \end{equation}
\end{corollary}
\begin{beweis}
  Note that $I(A:B) \geq S(A)$ if and only if $S(B) \geq S(C)$.
  Hence, we only have to show that for $\rho=\rho'$ with $S(B) < S(C)$
  and $S(B) < S(A)$, the lower bound (\ref{eq:C-ass:lower-3})
  in theorem~\ref{thm:C-ass:lower}
  is at least as large as $\frac{1}{2}S(A)$:
  \begin{equation*}\begin{split}
    &\left[ 1-\frac{S(B)}{S(A)} \right] S(C) + \frac{S(B)}{S(A)} S(B) \\
    &\phantom{===}
      \geq \left[ 1-\frac{S(B)}{S(B)+S(C)} \right] S(C) + \frac{S(B)}{S(B)+S(C)} S(B) \\
    &\phantom{===}
      =    \frac{S(B)^2+S(C)^2}{S(B)+S(C)}                                            \\
    &\phantom{===}
      \geq \frac{1}{2}\bigl[ S(B)+S(C) \bigr]   \geq \frac{1}{2}S(A).
  \end{split}\end{equation*}
  The first line comes from the subadditivity of entropy, $S(A) \leq S(B)+S(C)$,
  and using the assumption of $S(B) < S(C)$: substituting
  $S(B)+S(C)$ for $S(A)$ makes the weight of the smaller
  quantity smaller in the above convex combination.
  In the third line we use the arithmetic-geometric mean inequality,
  and subadditivity once more.
\end{beweis}

The result that $C_A^\rightarrow({\cal N}) \geq \frac{1}{2}\log d_A$
is somewhat reminiscent of an earlier observation by Fan~\cite{fan}:
that among the
``standard'' maximally entangled states in dimensions $d\times d$,
any set of $\leq \sqrt{2d}$ is LOCC-distinguishable with certainty.
Merging of quantum sources (lemma~\ref{lemma:merging}) gives is here an
improvement in the asymptotic setting:
consider any ensemble $\{ p_i,\varphi_i \}$ of orthogonal entangled states
on $BC$ such that $S(B) > S(BC)$ for the state $\rho = \sum_i p_i\varphi_i$.
(For example, less than $d$ maximally entangled states in
dimensions $d\times d$ with equal probabilities.)
Then, for sufficiently many independent samples from the ensemble,
Charlie can merge the unknown state from the ensemble
with Bob's (at least with high fidelity and for a large-probability
set of the ensemble), who then can distinguish them perfectly
as they are orthogonal.

Another important lower bound, that is actually better than the above
theorem and corollary for $d_A=2,3$, is proved in~\cite{hayden:king}:
$C_A^\rightarrow({\cal N}) \geq 1$ for every channel,
which settles the capacity question for qubit input system.

\section{An upper bound on the\protect\\ PPT-decoded classical capacity}
\label{sec:upperbound}
In this section we will prove a general upper bound on the PPT-decoded
classical capacity of a channel, and then demonstrate its usefulness by
analysing a class of examples, in the following section.

Before we embark on this, we note that Badzi\c{a}g et al.~\cite{badziag:etal}
have shown the following interesting bound:
\begin{proposition}[\cite{badziag:etal}, Thm.~1]
  \label{prop:badziag}
  Consider an ensemble of pure states
  $\ket{\varphi_i} \in {\cal H}_B \otimes {\cal H}_C$, with
  probabilities $p_i$, and an LOCC-implemented POVM $(M_j)_j$.
  Then, with the joint distribution $\Pr\{ X=i,Y=j\} = p_i \tr(\varphi_i M_j)$
  of random variables $X$ and $Y$,
  the Shannon mutual information is upper bounded as
  \begin{equation}
    \label{eq:badziag}
    I(X:Y) \leq S(\rho_B) + S(\rho_C) - \overline{E},
  \end{equation}
  where $\rho_{BC} = \sum_i p_i\varphi_i$ is the average
  state and $\rho_B$, $\rho_C$ are the reduced states,
  and $\overline{E} = \sum_i p_i E(\varphi_i)$ is
  the averaged pure state entanglement of the ensemble,
  and $E(\varphi) = S(\tr_C \varphi)$.
  \qed
\end{proposition}
This means that one obtains an upper bound on the ``locally (rather: LOCC)
accessible information''. An interesting feature is that the term
$\overline{E}$ vanishes if all states in the ensemble are
products, but then in the example of~\cite{nonlocal:w-o-ent} the
above inequality is not tight.
This motivates the conjecture that the above bound
may be true for a much wider class of POVMs including all
separable POVMs --- indeed, as we will see at the end of this
section, it holds true if the POVM is only PPT.

The following lemma is an adaptation of a result by
Owari and Hayashi~\cite{owari:hayashi},
whose is an elegant reformulation
and proof of an insight by Nathanson~\cite{nathanson},
to the case of (small) error in the detection, not quite maximal
entanglement, and PPT POVM elements:

\begin{lemma}
  \label{lemma:detector}
  Consider Hilbert spaces ${\cal H}_B$ and ${\cal H}_C$
  of dimensions $d_B \leq d_C$, respectively, and
  a pure state $\ket{\varphi} \in {\cal H}_B \otimes {\cal H}_C$ with
  $E(\varphi) \geq \log d_B - \Delta$.
  Then, for any PPT POVM element $M$ (i.e., $0\leq M\leq \1$ and
  $M^\Gamma \geq 0$), such that $\tr(\varphi M) \geq 1-\e$, and for every $K>1$,
  \begin{align}
    \label{eq:detector1}
    \tr M &\geq \Bigl( 1-\e-\sqrt{2}\sqrt[4]{\Delta}\; \Bigr) \; d_B, \\ 
    \label{eq:detector2}
    \tr M &\geq \left( 1-\e-\sqrt{\frac{\Delta+1}{\log K}} \right) \frac{d_B}{K}.
  \end{align}
  (The first bound is best for ``small'' $\Delta$, whereas the second
  will serve well in the regime of ``large'' $\Delta$.)
\end{lemma}
\begin{beweis}
  For eq.~(\ref{eq:detector1}) we observe that the condition
  $E(\varphi) = S(\tr_C \varphi) \geq \log d_B - \Delta$
  can be rewritten as
  $$D\left( \tr_C\varphi \Big\| \frac{1}{d_B}\1_B \right) \leq \Delta,$$
  hence by Pinsker's inequality (lemma~\ref{lemma:pinsker})
  $$\frac{1}{2}\left\| \tr_C\varphi - \frac{1}{d_B}\1_B \right\|_1 \leq \sqrt{\delta}.$$
  Hence, using lemmas~\ref{lemma:F-tracenorm} and~\ref{lemma:UJ},
  there exists a maximally entangled state
  $\widehat{\varphi}$ (i.e. with $d_B$ Schmidt coefficients $1/d_B$)
  such that $F(\varphi,\widehat{\varphi}) \geq (1-\sqrt{\delta})^2$,
  which implies (lemma~\ref{lemma:F-tracenorm} once more)
  $$\frac{1}{2}\left\| \varphi - \widehat{\varphi} \right\|_1 \leq \sqrt{2}\sqrt[4]{\delta}.$$
  From this get on the one hand
  $$\tr(\widehat{\varphi}M) \geq \tr(\varphi M) - \sqrt{2}\sqrt[4]{\delta}
                            \geq 1 - \e - \sqrt{2}\sqrt[4]{\delta}.$$
  On the other hand, using $M^\Gamma \geq 0$,
  \begin{equation*}\begin{split}
    \tr(\widehat{\varphi}M)
         &=    \tr\left( \widehat{\varphi}^\Gamma M^\Gamma \right)
          \leq \tr\left( \bigl| \widehat{\varphi}^\Gamma \bigr| M^\Gamma \right) \\
         &=    \tr\left( \frac{1}{d_B}\1 \, M^\Gamma \right)
          =    \frac{1}{d_B}\tr M^\Gamma
          =    \frac{1}{d_B} \tr M.
  \end{split}\end{equation*}
  Here, we have used the modulus of an operator,
  $|A| = \sqrt{A^* A}$, and the fact that for a maximally
  entangled state, the partial transpose is the (unitary!)
  swap operator, divided by the Schmidt rank.
  This concludes the proof of eq.~(\ref{eq:detector1}).

  For eq.~(\ref{eq:detector2}), let the Schmidt coefficients of $\varphi$
  be denoted $\lambda_j$ ($j=1,\ldots,d_B$), in decreasing order.
  We show first that
  \begin{equation}
    \label{eq:small-hills}
    q := \sum \{ \lambda_j : \lambda_j > K/d_B \} \leq \frac{\Delta+1}{\log K}.
  \end{equation}
  For this, assume that the first $L$ Schmidt coefficients $\lambda_j$
  exceed $K/d_B$. From monotonicity of $H$ under majorisation (see
  Alberti and Uhlmann~\cite{alberti:uhlmann})
  we see that the entropy of the distribution is maximised
  when $L = q\, d_B/K$ and the distribution has two flat sections:
  the first $L$ values are $q/L = K/d_B$, and the remaining
  $d_B-L$ values are $(1-q)/(d_B-L)$.
  (It is inessential for our argument that such $L$ may be
  non-integer: we only will overestimate the following entropy
  a little bit.)
  Now, this maximal entropy is
  \begin{equation*}\begin{split}
    H &=    H(q,1-q) + q\log L + (1-q)\log(d_B-L) \\
      &\geq E(\varphi) \geq \log d_B - \Delta.
  \end{split}\end{equation*}
  Rearranging this, using $H(q,1-q) \leq 1$,
  and substituting $L/d_B = q/K$, this finally yields
  \begin{equation*}\begin{split}
    \Delta+1 &\geq -q\log\frac{q}{K} -(1-q)\log\left(1-\frac{q}{K}\right) \\
             &=    q\log K -q\log q -(1-q)\log\left(1-\frac{q}{K}\right)  \\
             &\geq q\log K,
  \end{split}\end{equation*}
  as claimed.

  Now construct a pure state $\widetilde{\varphi}$ from $\varphi$
  by removing all Schmidt coefficients exceeding $K/d_B$ (and
  normalising such as to obtain a unit vector): it is straightforward
  to check that $F(\varphi,\widetilde{\varphi}) = 1-q$, with $q$
  taken from eq.~(\ref{eq:small-hills}), hence (by lemma~\ref{lemma:F-tracenorm})
  $$\frac{1}{2}\left\| \varphi - \widetilde{\varphi} \right\|_1 \leq \sqrt{q}.$$
  From here we can proceed much as before: we first get
  $$\tr(\widetilde{\varphi}M) \geq \tr(\varphi M) - \sqrt{q}
                              \geq 1 - \e - \sqrt{q}.$$
  On the other hand, using $M^\Gamma \geq 0$ once more,
  \begin{equation*}\begin{split}
    \tr(\widetilde{\varphi}M)
         &=    \tr\left( \widetilde{\varphi}^\Gamma M^\Gamma \right)
          \leq \tr\left( \bigl| \widetilde{\varphi}^\Gamma \bigr| M^\Gamma \right) \\
         &\leq \tr\left( \frac{K}{d_B}\1 \, M^\Gamma \right)
          =    \frac{K}{d_B}\tr M^\Gamma
          =    \frac{K}{d_B} \tr M,
  \end{split}\end{equation*}
  which concludes the proof of eq.~(\ref{eq:detector2}).
\end{beweis}

\begin{remark}
  \label{rem:markham-etal}
  {\rm
  On completing the present manuscript its author has become aware
  of the recent paper~\cite{markham}. It contains lower bounds similar to
  the above, for the error-free case, which are actually a bit better:
  for example, for a pure state $\varphi$ and separable/PPT POVM element $M$
  with $\tr(\varphi M) = 1$, it holds that $\log \tr M \geq E(\varphi)$.
  It will be interesting to follow the development of the elegant
  techniques of~\cite{markham} further, to deal with error probabilities.
  }
\end{remark}

\begin{theorem}
  \label{thm:upperbound}
  Let $(\varphi_i,M_i)_{i=1}^N$ be a code of pure states
  $\ket{\varphi_i} \in {\cal H}_B\otimes{\cal H}_C$, such that
  for all $i$, $E(\varphi_i) \geq \log d_B - \delta$,
  and PPT POVM $(M_i)_{i=1}^N$, with error probability $\leq \e$.
  Then, if $\e+\sqrt{2}\sqrt[4]{\d} < 1$,
  $$N \leq \Bigl( 1-\e-\sqrt{2}\sqrt[4]{\delta}\; \Bigr)^{-1} d_C.$$
\end{theorem}
\begin{beweis}
  Since by assumption all of the operators $M_i$ are PPT, we can
  use eq.~(\ref{eq:detector1}) of lemma~\ref{lemma:detector}:
  $$\text{for all }i,\ 
    \tr M_i \geq \Bigl( 1-\e-\sqrt{2}\sqrt[4]{\delta}\; \Bigr) \; d_B.$$
  On the other hand, from the POVM condition that
  $\sum_{i=1}^N M_i \leq \1_{BC}$, we get that
  $\sum_{i=1}^N \tr M_i \leq d_B d_C$, which yields the
  upper bound on $N$ as advertised.
\end{beweis}

\begin{theorem}
  \label{thm:upperbound2}
  Let $(\varphi_i,M_i)_{i=1}^N$ be a code of pure states
  $\ket{\varphi_i} \in {\cal H}_B\otimes{\cal H}_C$, such that
  for all $i$, $E(\varphi_i) \geq \log d_B - \Delta_i$,
  and PPT POVM $(M_i)_{i=1}^N$, with error probability $\leq \e$.
  Then, for $\gamma > 1/(1-\e)^2$,
  \begin{equation*}\begin{split}
    N &\leq \left( 1-\e-\sqrt{\frac{1}{\gamma}} \right)^{-1}
             \left( \frac{1}{N}\sum_{i=1}^N 2^{-\gamma(\Delta_i+1)} \right)^{-1} d_C \\
      &\leq \left( 1-\e-\sqrt{\frac{1}{\gamma}} \right)^{-1}
             2^{\sum_{i=1}^N \gamma(\Delta_i+1)/N} d_C.
  \end{split}\end{equation*}
\end{theorem}
\begin{beweis}
  Since by assumption all of the operators $M_i$ are PPT, we can
  use eq.~(\ref{eq:detector2}) of lemma~\ref{lemma:detector}:
  for the pair $(\varphi_i,M_i)$, we set $K_i = 2^{\gamma(\Delta_i+1)}$
  and obtain
  $$\text{for all }i,\ 
    \tr M_i \geq \left( 1-\e-\sqrt{\frac{1}{\gamma}} \right) 2^{-\gamma(\Delta_i+1)} d_B.$$
  On the other hand, from the POVM condition that
  $\sum_{i=1}^N M_i \leq \1_{BC}$, we get that
  $\sum_{i=1}^N \tr M_i \leq d_B d_C$, which yields the
  upper bound on $N$ as claimed; for the final upper bound we
  have to use the arithmetic-geometric mean inequality.
\end{beweis}

\begin{corollary}
  \label{cor:capacity-upperbound}
  Let $U: {\cal H}_A \longrightarrow {\cal H}_B \otimes {\cal H}_C$
  ($d_B \leq d_C$ without loss of generality), and assume for
  the subspace ${\cal S} = U{\cal H}_A \subset {\cal H}_B \otimes {\cal H}_C$,
  that for all $n$ and for all $\ket{\varphi} \in {\cal S}^{\otimes n}$,
  $E(\varphi) \geq n(\log d_B - \delta)$.
  Then,
  $$C_A^{\rm ppt}(U) \leq \log d_C + \delta.$$
\end{corollary}
\begin{beweis}
  For a given blocklength $n$, consider a PPT-decoded code
  $(\varphi_i,M_i)_{i=1}^N$ of error $\leq \e$
  and rate $R=\frac{1}{n}\log N$.

  We now use the previous theorem~\ref{thm:upperbound2} with
  local dimensions $d_B^n$ and $d_C^n$, and $\Delta_i = n\delta$.
  This yields, for $\gamma > 1/(1-\e)^2$,
  $$N \leq \left( 1 - \e - \sqrt{\frac{1}{\gamma}} \right)^{-1}
           2^{\gamma(1+n\delta)} d_C^n.$$
  For the rate this means
  $$R \leq \log d_C + \gamma\delta + O\left(\frac{1}{n}\right),$$
  and since in the limit $n\rightarrow\infty$, $\e\rightarrow 0$
  we can choose $\gamma$ arbitrarily close to $1$, every
  asymptotically achievable rate is bounded above
  by $\log d_C + \delta$, as claimed.
\end{beweis}

\begin{remark}
  \label{rem:superadd}
  {\rm
  The assumption of corollary~\ref{cor:capacity-upperbound} is widely believed
  to actually follow from the case $n=1$.
  This is known as the superadditivity conjecture
  for the entanglement of formation~\cite{shor:add}:
  \begin{equation}
    \label{eq:superadd}
    E_F(\rho_{B_1B_2 C_1C_2}) \geq E_F(\rho_{B_1C_1}) + E_F(\rho_{B_2C_2}),
  \end{equation}
  where the entanglement of formation, $E_F$~\cite{BDSW}, is the convex
  hull of the reduced state entropy function $E$.

  Note that by assumption of $E(\varphi) \geq \log d_B - \delta$ for all
  $\ket{\varphi} \in {\cal S}$, every state $\rho$ supported on ${\cal S}$
  has $E_F(\rho) \geq \log d_B - \delta$, hence by induction
  on $n$ we get $E(\varphi) \geq n(\log d_B - \delta)$
  for all $\ket{\varphi}\in{\cal S}^{\otimes n}$.
  }
\end{remark}

\begin{remark}
  \label{rem:badziag-ppt}
  {\rm
  It should be obvious that the bound of Badzi\c{a}g et al.~\cite{badziag:etal},
  stated above as proposition~\ref{prop:badziag}, implies the bound
  of corollary~\ref{cor:capacity-upperbound} for the \emph{LOCC-assisted
  capacity}:
  $$C_A^\leftrightarrow({\cal N}) \leq \log d_C + \delta.$$
  Now we want to show that our theorems for PPT-decoders imply that
  the inequality~(\ref{eq:badziag}) holds if $(M_j)_j$ is a PPT POVM.
  }
\end{remark}
\begin{beweis}[(Sketch)]
  As before, the ensemble and the POVM give us random variables
  with joint distribution
  $$\Pr\{ X=i,Y=j \} = p_i \tr(\varphi_i M_j).$$
  Now, by Shannon's channel coding theorem~\cite{shannon}, random
  coding on large block length $n$ gives, with high probability,
  a good code ${\cal C}$ of rate achieving $I(X:Y)$. In fact, since the codewords
  $I=i_1\ldots i_n$ are chosen at random according to the distribution
  $p_I=p_{i_1}\cdots p_{i_n}$, most codewords will be typical, i.e.,
  each letter $i$ occurs $\approx np_i$ times. Expurgating the untypical
  codewords we loose no rate asymptotically, but now all codewords
  can be assumed to be typical.

  So, we have, for arbitrary $\eta>0$ and
  for all sufficiently large $n$, a PPT-decoded code
  $(\Phi_I,D_I)_{I\in{\cal C}}$, with
  $$\ket{\Phi_I} = \ket{\varphi_{i_1}} \otimes \cdots \otimes \ket{\varphi_{i_n}}$$
  and PPT operators $D_I$, such that
  \begin{align*}
                          \frac{1}{n}\log|{\cal C}| &\geq I(X:Y) - \eta, \\
    \forall I\in{\cal C}\ \frac{1}{n}E(\Phi_I)      &\geq \overline{E} - \eta,
  \end{align*}
  and error probability $\e\rightarrow 0$ as $n\rightarrow\infty$.

  Now we can further modify the code by using the typical subspace
  projectors $\Pi_B$ and $\Pi_C$ of $\rho_B^{\otimes n}$ and
  $\rho_C^{\otimes n}$, respectively~\cite{quantum:coding}:
  create a new POVM (now on the tensor product of the two
  typical subspaces) with operators
  $$D_I' := (\Pi_B\otimes\Pi_C) D_I (\Pi_B\otimes\Pi_C),$$
  which is easily seen to inherit the PPT property from $(D_I)_I$.
  On the other hand (see~\cite{winter:qstrong}) this degrades the
  error probability only marginally, say increasing it to
  $2\e$, and the local dimensions of Bob and Charlie are now
  bounded by $2^{n(S(\rho_B)+\eta)}$ and $2^{n(S(\rho_C)+\eta)}$.

  At this point we can finish, invoking theorem~\ref{thm:upperbound2}:
  \begin{equation*}\begin{split}
    I(X:Y) - \eta &\leq \frac{1}{n}\log|{\cal C}|                                       \\
                  &\!\!\!\!\!\!\!\!\!\!\!\!\!\!\!
                   \leq S(\rho_C) + \eta
                        + \gamma\bigl( S(\rho_B)+\eta-\overline{E}+\eta \bigr) + O(1/n) \\
                  &\!\!\!\!\!\!\!\!\!\!\!\!\!\!\!
                   \leq S(\rho_B)+S(\rho_C)-\overline{E}                                \\
                  &\!\!\!\!\!\!
                        + (\gamma-1)\log d_B + (2\gamma+1)\eta + O(1/n).
  \end{split}\end{equation*}
  Since $\eta>0$ was arbitrary and also $\gamma>1$ is arbitrary as
  the error probability $\e\rightarrow 0$ and $n\rightarrow\infty$,
  we obtain the desired bound.
\end{beweis}

\section{An example almost\protect\\ meeting the lower bound~(\ref{eq:C-ass:lower2})%
\protect\\ \ldots modulo additivity conjecture}
\label{sec:example}
It is clear that the upper bounds on $C_A^{\rm ppt}$ developed in
the previous section are not very tight in general. In particular, for the
bound of corollary~\ref{cor:capacity-upperbound} to be nontrivial,
the dimension $d_A$ of the subspace ${\cal S}$ must be significantly
larger than $d_C$ ($\geq d_B$).

Fortunately, we can use here the recently discovered existence of
quite large subspaces in $d_B \times d_C$ which meet the requirements
of theorem~\ref{thm:upperbound} and, assuming the universal
validity of $E_F$-superadditivity~(\ref{eq:superadd}),
of corollary~\ref{cor:capacity-upperbound}:

\begin{proposition}[\cite{generic}, Thm.~IV.1]
  \label{prop:subspace}
  Let ${\cal H}_B$ and ${\cal H}_C$ be quantum systems of dimension
  $d_B$ and $d_C$, respectively, for $d_C \geq d_B \geq 3$.
  Let $0 < \alpha < \log d_B$.
  Then there exists a subspace ${\cal S} \subset {\cal H}_B \otimes {\cal H}_C$
  of dimension
  \begin{equation*}
    \left\lfloor d_B d_C \frac{\G\a^{2.5}}{(\log d_B)^{2.5}} \right\rfloor
  \end{equation*}
  such that all states $\ket{\varphi} \in {\cal S}$ have entanglement at least
  \begin{equation*}
    E(\varphi) = S(\ph_A) \geq \log d_B - \frac{1}{\ln 2}\frac{d_B}{d_C} - \alpha,
  \end{equation*}
  where $\G$ is an absolute constant which may
  be chosen to be $1/1753$. \qed
\end{proposition}

With $d_B=d_C=d$ and $\alpha = 20$
we are thus guaranteed a subspace ${\cal S} \subset {\cal H}_B \otimes {\cal H}_C$
of dimension $d_A = \left\lfloor d^2 \frac{1.0204}{(\log d)^{2.5}} \right\rfloor$,
such that all states $\ket{\varphi}\in{\cal S}$ have
entanglement $E(\varphi) \geq \log d - 21.5$.
The channel ${\cal N}$ we now consider is simply the
embedding $U$ of ${\cal H}_A = {\cal S}$ into the
tensor product, followed by a partial trace over $C$.
Of course this makes nontrivial sense only for rather large $d$
(namely $d\geq 128$, when $d_A$ starts becoming larger than $d$),
which we silently assume from here on.

As mentioned a couple of times already, we will now assume
the superadditivity conjecture (remark~\ref{rem:superadd}),
which means that we will
proceed under the assumption that
$$\text{for all }\ket{\varphi}\in{\cal S}^{\otimes n},\ E(\varphi) \geq n(\log d - 21.5).$$
Then corollary~\ref{cor:capacity-upperbound} gives us the bound
$$C_A^{\rm ppt}(U) \leq \log d + 21.5
                   \leq \frac{1}{2}\log d_A + 2.5 \log \log d_A + 27.$$
The point here being that this comes close to the lower bound
of theorem~\ref{thm:C-ass:lower}, up to a doubly logarithmic term
and a (rather large) constant.

Finally, let us mention that using proposition~\ref{prop:subspace}
we can also produce an example catering to theorem~\ref{thm:upperbound}:
simply choose $d_B=d$, $d_C = \frac{2}{\delta\ln 2} d$ and $\alpha = \d/2$.

\section{Discussion}
\label{sec:discussion}
We have shown some new lower and upper bounds on environment-assisted
and PPT decoded capacities of quantum channels.
In particular, we have shown that the environment-assisted classical capacity
is always at least half the input bandwidth, and we have exhibited
a class of examples which indicate that this factor of
$1/2$ is indeed attained in the worst case, even when the broader
class of PPT decodings is permitted.
This seems quite remarkable, as the lower bound is actually
achieved some of the time by transmitting \emph{quantum}
information from Alice to Bob, and part of the time by
transmitting quantum information partly to Charlie and partly
to Bob (all of course with one-way LOCC help of Charlie to Bob).

In the process we have generalised a previously known bound on the
locally accessible information to PPT POVMs;
perhaps this will help clarifying the conceptual origin
of such bounds (which in~\cite{badziag:etal} is proved by
going through a general LOCC protocol).
It is however quite clear by simple examples
that this upper bound cannot be optimal
in general, even asymptotically and with coding;
see also~\cite{more-nonlocal:less-ent} which indicates that
the upper bound cannot be in terms of local
entropies and entanglement alone.

Our work still leaves wide open the problem of finding
a formula for the
assisted classical capacities $C_A^\rightarrow$ and $C_A^\leftrightarrow$.
It seems that the main advance to be made is in trying to
tighten the upper bounds on the locally accessible information.
And of course we would like to narrow the gap between the lower bound
and the worst-case upper bound for $C_A^\rightarrow$ and $C_A^{\rm ppt}$,
and preferably so without resorting to unproven conjectures.

\acknowledgments
Conversations with Chris King
on the topics covered in this paper are gratefully acknowledged.
The author is supported by the EU project RESQ (contract
no.~IST-2001-37559), by the U.K. Engineering and Physical
Sciences Research Council's ``IRC QIP'', and a
University of Bristol Research Fellowship.

\appendix

\section{Technical results}

\begin{lemma}[See~\cite{fuchs:vandegraaf}]
  \label{lemma:F-tracenorm}
  For two mixed states $\rho$, $\sigma$, the fidelity is
  $F(\rho,\sigma) = \| \sqrt{\rho}\sqrt{\sigma} \|_1^2
   = \left( \tr\!\sqrt{\!\sqrt{\rho}\sigma\sqrt{\rho}} \right)^2$,
  with the trace norm $\| A \|_1 = \tr\sqrt{A^*A}$. Then,
  $$1-\sqrt{F(\rho,\sigma)} \leq \frac{1}{2}\| \rho-\sigma \|_1
                                     \leq \sqrt{1-F(\rho,\sigma)}.$$
\end{lemma}

\begin{lemma}[See~\cite{uhlmann:jozsa}]
  \label{lemma:UJ}
  Let $\rho$, $\sigma$ be states on ${\cal H}$ and let
  $\ket{\varphi},\ket{\psi} \in {\cal H}\otimes{\cal H}$
  vary over purifications of $\rho,\sigma$, respectively.
  Then,
  $$F(\rho,\sigma) = \max_{\varphi,\psi} F(\varphi,\psi).$$
  Observe that for pure states,
  $F(\varphi,\psi) = \tr\varphi\psi = | \langle \varphi \ket{\psi} |^2$.
\end{lemma}

\begin{lemma}[Pinsker's inequality, see~\cite{ohya:petz}]
  \label{lemma:pinsker}
  For two arbitrary states $\rho$, $\sigma$, the relative entropy
  is defined as $D(\rho \| \sigma) = \tr\bigl( \rho(\log\rho-\log\sigma)\bigr)$
  [which may be $+\infty$ if the support of $\rho$ is not contained
  in that of $\sigma$].
  Then,
  $$\left( \frac{1}{2}\| \rho-\sigma \|_1 \right)^2 \leq D(\rho\|\sigma).$$
\end{lemma}


\begin{thebibliography}{99}


  \bibitem{alberti:uhlmann} P. M. Alberti, A. Uhlmann, \emph{Stochasticity and partial
    order: doubly stochastic maps and unitary mixing},
    Kluwer: Dordrecht-Boston, 1982.

  \bibitem{badziag:etal} P. Badzi\c{a}g, M. Horodecki, A. Sen(De), U. Sen,
    ``Locally Accessible Information: How Much Can the Parties Gain by
    Cooperating?'', Phys. Rev. Lett., vol. 91, no. 11, 117901, 2003.
    S. Ghosh, P. Joag, G. Kar, S. Kunkri, A. Roy, ``Locally accessible information
    and distillation of entanglement'', Phys. Rev. A, vol. 71, no. 2,
    012321, 2005.


  \bibitem{BDSW} C. H. Bennett, D. P. DiVincenzo, J. A. Smolin, W. K. Wootters,
    ``Mixed-state entanglement and quantum error correction'', Phys. Rev. A,
    vol. 54, no. 5, pp. 3824-3851, 1996.

  \bibitem{nonlocal:w-o-ent} C. H. Bennett, D. P. DiVincenzo, C. A. Fuchs, T. Mor,
    E. Rains, P. W. Shor, J. A. Smolin, W. K. Wootters, ``Quantum nonlocality
    without entanglement'', Phys. Rev. A, vol. 59, no. 2, 1070-1091, 1999.


  \bibitem{chefles} A. Chefles, ``Condition for unambiguous state discrimination
    using local operations and classical communication'', Phys. Rev. A,
    vol. 69, no. 5, 050307(R), 2004.



  \bibitem{devetak:Q} I. Devetak, ``The Private Classical Capacity and Quantum
    Capacity of a Quantum Channel'',
    IEEE Trans. Inf. Theory, vol. 51, no. 1, pp. 44-55, 2005.





  \bibitem{fan} H. Fan, ``Distinguishability and Indistinguishability by Local
    Operations and Classical Communication'', Phys. Rev. Lett., vol. 92,
    no. 17, 177905, 2004.


  \bibitem{fuchs:vandegraaf} C. A. Fuchs, J. van de Graaf, ``Cryptographic
    Distinguishability Measures for Quantum-Mechanical States'',
    IEEE Trans. Inf. Theory, vol. 45, no. 4, pp. 1216-1227, 1999.

  \bibitem{ghosh:etal} S. Ghosh, G. Kar, A. Roy, A. Sen(De), U. Sen,
    ``Distinguishability of Bell States'', Phys. Rev. Lett., vol. 87,
    no. 27, 277902, 2001.
    S. Ghosh, G. Kar, A. Roy, D. Sarkar, ``Distinguishability
    of maximally entangled states'', Phys. Rev. A, vol. 70, no. 2, 022304, 2004.

  \bibitem{lost:found} M. Gregoratti, R. F. Werner, ``Quantum Lost and Found'',
    J. Mod. Optics, vol. 50, no. 6\&7, pp. 913-933, 2003.

  \bibitem{more-lost-and-found} M. Gregoratti, R. F. Werner, ``On quantum error
    correction by classical feedback in discrete time'', e-print
    {\tt quant-ph/0403092}, 2004.


  \bibitem{markham} M. Hayashi, D. Markham, M. Murao, M. Owari, S. Virmani,
    ``LOCC State Discrimination and Multipartite Entanglement Measures'',
    e-print {\tt quant-ph/0506170}, 2005.


  \bibitem{hayden:king} P. Hayden, C. King, ``Correcting quantum channels
    by measuring the environment'', e-print {\tt quant-ph/0409026}, 2004.

  \bibitem{generic} P. Hayden, D. W. Leung, A. Winter, ``Aspects of
    generic entanglement'', e-print {\tt quant-ph/0407049}, 2004.



  \bibitem{more-nonlocal:less-ent} M. Horodecki, A. Sen(De), U. Sen, K.
    Horodecki, ``Local Indistinguishability: More Nonlocality with Less Entanglement''
    Phys. Rev. Lett., vol. 90, no. 4, 047902, 2003.

  \bibitem{HOW} M. Horodecki, J. Oppenheim, A. Winter, ``Quantum
    information can be negative'', e-print {\tt quant-ph/0505062}, 2005.
    As ``Partial quantum information'' to appear in Nature.







  \bibitem{nathanson} M. Nathanson, ``Distinguishing Bipartitite Orthogonal
    States using LOCC: Best and Worst Cases'', e-print {\tt quant-ph/0411110}, 2004.

  \bibitem{ohya:petz} M. Ohya, D. Petz, \emph{Quantum Entropy and Its Use},
    Springer Verlag, Berlin, 1993.

  \bibitem{owari:hayashi} M. Owari, M, Hayashi, ``Local copying of orthogonal
    maximally entangled states and its relation to local discrimination'',
    e-print {\tt quant-ph/0411143}, 2004.


  \bibitem{rains} E. M. Rains, ``A Semidefinite Program for Distillable Entanglement'',
    IEEE Trans. Inf. Theory, vol. 47, no. 7, pp. 2921-2933, 2001.



  \bibitem{quantum:coding} B. Schumacher, ``Quantum Coding'',
    Phys. Rev. A, vol. 51, no. 4, pp. 2738-2747, 1995.
    R. Jozsa, B. Schumacher, ``A new proof of the quantum noiseless coding
    theorem'', J. Mod. Optics, vol. 41, no. 12, pp. 2343-2349, 1994.



  \bibitem{shor:add} K. G. H. Vollbrecht, R. F. Werner, ``Entanglement measures
    under symmetry'', Phys. Rev. A, vol. 64, no. 6, 062307, 2001.
    K. Matsumoto, T. Shimono, A. Winter, ``Remarks on Additivity of the Holevo
    Channel Capacity and of the Entanglement of Formation'', Comm. Math. Phys.,
    vol. 246, no. 3, pp. 427-442, 2004.
    P. W. Shor, ``Equivalence of Additivity Questions in Quantum
    Information Theory'', Comm. Math. Phys., vol. 246, no. 3, pp. 453-472, 2004.
    A. A. Pomeransky, ``Strong superadditivity of the entanglement of formation
    follows from its additivity'', Phys. Rev. A, vol. 68, no. 3, 032317, 2003.

  \bibitem{shannon} C. E. Shannon, ``A mathematical theory of communication'',
    Bell. Syst. Tech. J., vol. 27, pp. 379-423 and 623-656, 1948.

  \bibitem{EoA:oho} J. A. Smolin, F. Verstraete, A. Winter, ``Entanglement of
    assistance and multipartite state distillation'',
    e-print {\tt quant-ph/0505038}, 2005.

  \bibitem{stinespring} W. F. Stinespring, ``Positive functions on C${}^*$-algebras'',
    Proc. Amer. Math. Soc., vol. 6, pp. 211-216, 1955.



  \bibitem{uhlmann:jozsa} A. Uhlmann, ``The `transition probability' in the state space
    of a $*$-algebra'', Rep. Math. Phys., vol. 9, pp. 273-279, 1976.
    R. Jozsa, ``Fidelity for mixed quantum states'', J. Mod. Optics,
    vol. 41, no. 12, pp. 2315-2323, 1994.



  \bibitem{walgate:etal} J. Walgate, A. Short, L. Hardy, V. Vedral, ``Local
    Distinguishability of Multipartite Orthogonal Quantum States,
    Phys. Rev. Lett., vol. 85, no. 23, pp. 4972-4975, 2000.

  \bibitem{walgate:hardy} J. Walgate, L. Hardy, ``Nonlocality, Asymmetry, and
    Distinguishing Bipartite States'', Phys. Rev. Lett., vol. 89, no. 14,
    147901, 2002.

  \bibitem{winter:qstrong} A. Winter, ``Coding theorem and strong converse for
    quantum channels'', IEEE Trans. Inf. Theory, vol. 45, no. 7, pp. 2481-2485, 1999.


\end{thebibliography}
\end{document}